\documentstyle[prl,aps,twocolumn]{revtex} 

\newcommand{\da}{^{\dagger}}
\newcommand{\zo}{^{(0)}}
\newcommand{\nonpsi}{\tilde{\psi}}
\newcommand{\r}{{\bf r}}

\newcommand{\vect}[1]{{\mathbf{#1}}}
\newcommand{\parenth}[1]{\left( #1 \right)}

\newcommand{\angles}[1]{\langle#1\rangle}

\newcommand{\partder}[2]{\frac{\partial{#1}}{\partial{#2}}}
\newcommand{\der}[2]{\frac{d{#1}}{d{#2}}}

\newcommand{\be}{\begin{equation}}
\newcommand{\ee}{\end{equation}}
\newcommand{\ba}{\begin{eqnarray}}
\newcommand{\ea}{\end{eqnarray}}
\newcommand{\bastar}{\begin{eqnarray*}}
\newcommand{\eastar}{\end{eqnarray*}}
\newcommand{\la}{\label}

\begin{document}
\draft
\title{Adiabaticity Criterion for Moving Vortices in Dilute 
Bose-Einstein Condensates}

\author{S. M. M. Virtanen, T. P. Simula, and M.~M.~Salomaa}
\address{Materials Physics Laboratory, 
Helsinki University of Technology\\
P.~O.~Box 2200 (Technical Physics), FIN-02015 HUT, Finland}
\date{\today}

\maketitle
\begin{abstract}
Considering a moving vortex line in a dilute atomic Bose-Einstein
condensate within time-dependent Hartree-Fock-Bogoliubov-Popov 
theory, we derive a criterion for the quasiparticle excitations
to follow the vortex core rigidly. The assumption of adiabaticity,
which is crucial for the validity of the stationary self-consistent
theories in describing such time-dependent phenomena, is shown to
imply a stringent criterion for the velocity of the vortex line.
Furthermore, this condition is shown to be violated in
the recent vortex precession experiments.
\end{abstract}
\hspace{5mm}


Since the first experimental realizations of Bose-Einstein condensation
in dilute, harmonically trapped atomic gases \cite{first_exp}, there has 
been great interest to investigate the superfluid properties of these 
unique quantum fluids. Due to the inherent connections between quantized
vorticity and superfluidity, this interest culminated as the
creation of vortices in trapped
condensates was demonstrated \cite{vortex_exp}. 
The recent experimental advances in manipulating vortices and observing
their dynamics are providing efficient tools to study the physics
of these interacting many-particle systems and to relate it to
the quantitative predictions of thermal field theories.

The structure and, in particular, the stability of vortices in
dilute Bose-Einstein condensates (BECs) has 
been under an extensive theoretical analysis 
\cite{fetter_review}. The majority of the studies has been carried 
out within the zero-temperature mean-field formalism consisting of the 
Gross-Pitaevskii (GP) and Bogoliubov equations. Within the
Bogoliubov approximation (BA), the excitation spectra of vortex states 
in statically trapped condensates have been shown to contain at least 
one mode with positive norm but negative energy \cite{neg_modes}. 
These anomalous modes have
crucial consequences for the superfluid properties of the condensates,
since they imply the vortices to be energetically unstable in nonrotating
traps. Furthermore, these states have been shown to manifest themselves
in the precession of the vortex line about the symmetry axis of the trap, 
with the precession frequency and direction determined by the 
excitation energy---especially, the negative energies
imply precession in the direction of the condensate flow 
\cite{prec_lcls}.

The predictions of the Bogoliubov approximation agree well with 
the experiments. The critical trap rotation frequencies for vortex
nucleation can be understood theoretically to good 
accuracy \cite{feder}. Also, the precession of vortices predicted by the 
GP equation has been experimentally observed \cite{new_exp}. 
The precession frequency and, in particular, its direction 
are in line with expectations based on the BA. In general, 
the mean-field theory has turned out to be remarkably successful 
in describing trapped BECs, 
including the vortex states and their dynamics \cite{dalfovo_review}.

However, the situation changes when the analysis is taken beyond 
the zero-temperature BA by self-consistently including the effects 
of the thermal gas component. Stationary self-consistent solutions for 
vortex states within the Popov approximation (PA) and its recently
proposed extensions contain no anomalous modes even in the
zero-temperature limit \cite{isoshima,our_letter}. 
This is due to the partial filling of the vortex core 
with the noncondensate, which serves to lift the anomalous quasiparticle
states to positive energies. The positive precession mode energies,
in turn, imply vortex precession opposite to the condensate flow, 
in evident contradiction with the experimental 
observations and the predictions of the BA.
In the light of the success of the self-consistent approximations
in predicting excitation spectra for irrotational condensates
\cite{dodd,hutchinson}, this discrepancy is surprising. 
In addition, the close agreement of the
BA with the results of the vortex precession experiments implies
that the mean-field approximation itself 
is not the cause of the failure of the PA. 

We suggest that the apparent disagreement between the PA and the experiments 
could be due to incomplete thermalization and/or inadequacy of 
the quasi-static formalism in describing 
moving vortices. In order to clarify the latter possibility, we show in
this paper that the validity of the quasi-static self-consistent mean-field
treatment in modelling moving vortices imposes for the vortex velocity
a stringent criterion, which seems to be violated in the precession
observations reported so far. This implies that at the 
observed velocities the quasiparticles can not follow the vortex
core rigidly, and its structure and spectrum are deformed from those 
of a static vortex.

In order to describe the dynamics of trapped BECs,
we use a time-dependent mean-field formalism based on the Popov
approximation \cite{tformalism}. 
Working in the grand-canonical formalism,
we start with the Heisenberg equation of motion
\be
\la{heisenberg}
i\hbar\partder{}{t}\psi(\r,t)
={\mathcal H}_0(\r)\psi(\r,t)+g\psi\da(\r,t)\psi(\r,t)\psi(\r,t)
\ee
for the field operator $\psi(\r,t)$ of a dilute boson gas. Above,
${\mathcal H}_0(\r)\equiv -\hbar^2\nabla^2/2m+V_{\rm tr}(\r)-\mu$
is the grand-canonical one-particle Hamiltonian corresponding to the
trapping potential $V_{\rm tr}(\r)$ and the chemical potential $\mu$,
and the coupling constant $g$ is related to the $s$-wave scattering
length $a$ by $g=4\pi\hbar^2a/m$. Inserting into the nonequilibrium
average of Eq.\ (\ref{heisenberg}) the Bogoliubov decomposition 
\be
\psi(\r,t)=\Phi(\r,t)+\nonpsi(\r,t)
\ee
of the field operator in terms of the $c$-number condensate wavefunction
$\Phi(\r,t)=\angles{\psi(\r,t)}$ and the noncondensate field operator 
$\nonpsi(\r,t)$, and treating the expectation values of noncondensate
operator products according to the Popov mean-field scheme,
we arrive at the generalized GP equation 
\be
i\hbar\partder{}{t}\Phi(\r,t)={\mathcal L}(\r,t)\Phi(\r,t)
-gn_{\rm c}(\r,t)\Phi(\r,t)
\ee
for the condensate wavefunction. Above,
${\mathcal L}(\r,t)\equiv {\mathcal H}_0(\r)+2gn(\r,t)$, and
\begin{mathletters}
\ba
n_{\rm c}(\r,t)&=&|\Phi(\r,t)|^2,\\
\tilde{n}(\r,t)&=&\angles{\nonpsi\da(\r,t)\nonpsi(\r,t)},\\
n(\r,t)&=&n_{\rm c}(\r,t)+\tilde{n}(\r,t)
\ea
\end{mathletters}
denote the condensate, noncondensate, and total densities, respectively.
Correspondingly, within the Popov mean-field approximation, one finds 
\be
\la{noncondheis}
i\hbar\partder{}{t}\nonpsi(\r,t)={\mathcal L}(\r,t)\nonpsi(\r,t)
+g\Phi^2(\r,t)\nonpsi\da(\r,t)
\ee
for the equation of motion of the noncondensate field operator. Substituting
into Eq.\ (\ref{noncondheis}) the Bogoliubov transformation
\be
\la{bogo}
\nonpsi(\r,t)=\sum_n[u_n(\r,t)\alpha_n-v_n^*(\r,t)\alpha\da_n]
\ee
of the field operator 
in terms of the bosonic quasiparticle operators $\alpha_n$ and $\alpha\da_n$,
we find that the quasiparticle amplitudes $u_n(\r,t)$ and $v_n(\r,t)$ satisfy
the time-dependent Hartree-Fock-Bogoliubov-Popov (TDHFBP) equations
\begin{mathletters}
\ba
i\hbar\partder{}{t}u_n(\r,t)&=&{\mathcal L}(\r,t)u_n(\r,t)
-g\Phi^2(\r,t)v_n(\r,t)\\
-i\hbar\partder{}{t}v_n(\r,t)&=&{\mathcal L}(\r,t)v_n(\r,t)
-g{\Phi^*}^2(\r,t)u_n(\r,t).
\ea 
\end{mathletters}
Introducing the matrix notations 
\be
f_n(\r,t)=\pmatrix{u_n(\r,t) \cr v_n(\r,t)},
\ee
\be
{\mathcal O}(\r,t)=\pmatrix{{\mathcal L}(\r,t) & -g\Phi^2(\r,t) 
\cr g{\Phi^*}^2(\r,t) & -{\mathcal L}(\r,t)},
\ee
the quasiparticle equations can be expressed in the compact form
\be
\la{TDHFBP}
i\hbar\partder{}{t}f_n(\r,t)={\mathcal O}(\r,t)f_n(\r,t).
\ee
For convenience, we also define positive- and negative-sign scalar products
of quasiparticle states by setting
\be
{\angles{f_i|f_j}}_{\pm}\equiv \int d\r\,[u_i^*(\r)u_j(\r)\pm
v_i^*(\r)v_j(\r)];
\ee
here and henceforth, we suppress the arguments of
functions when they are not needed for clarity. 
The requirement that the quasiparticle
operators $\alpha_n$, $\alpha\da_n$ satisfy canonical bosonic commutation
relations implies for the quasiparticle states the normalization
${\angles{f_i|f_j}}_- = \delta_{ij}$;
correspondingly, only states with positive norm 
are to be included in the Bogoliubov transformation of 
Eq.\ (\ref{bogo}). This normalization can be
straightforwardly verified to be consistent with the TDHFBP equations.

In case the mean fields and, hence, the operator ${\mathcal O}(\r,t)$ vary
slowly in time, we expect that the solutions of the TDHFBP equations
may be approximated by solving at each instant of time the corresponding
quasi-stationary eigenequations
\be
\la{bdgadiab}
E_n(t)f_n^{(0)}(\r,t)={\mathcal O}(\r,t)f_n^{(0)}(\r,t).
\ee
This adiabatic approximation is accurate if the transition rates,
as determined by the exact time development, of the quasi-stationary
states to each other are negligible. In order to formulate this
criterion quantitatively, we follow the treatment of Ref.\ \cite{schiff}.
Let $\{f\zo_i\}$ be a complete set of solutions of Eq.\ (\ref{bdgadiab});
especially, it contains the zero-energy solution
$f\zo_0\propto(\Phi_0,\Phi_0^*)^T$, where $\Phi_0$ is the solution of
the stationary GP equation. We orthonormalize the solutions by requiring
\begin{mathletters}
\la{orthon}
\be
|{\angles{f\zo_i|f\zo_j}}_-|=\delta_{ij}\quad (i\neq 0);\qquad 
{\angles{f\zo_0|f\zo_0}}_+=1.
\ee
In addition, we may impose the condition \cite{morgan}
\be
{\angles{f\zo_0|f\zo_i}}_+=0\quad (i\neq 0).
\ee
\end{mathletters}
In order to analyse the transitions of the 
quasi-stationary states to each other,
we expand the solutions of the TDHFBP equation in terms of them. Substitution
of the ansatz
\be
f_n(\r,t)=\sum_j a_{nj}(t)f\zo_j(\r,t)
e^{-\frac{i}{\hbar}\int_0^t E_j(t')dt'}
\ee
into Eq.\ (\ref{TDHFBP}) yields the coupled differential equations
\be
\sum_j [\dot{a}_{nj}f\zo_j+a_{nj}\dot{f}\zo_j]
e^{-\frac{i}{\hbar}\int_0^t E_j(t')dt'}=0,
\ee
where the dots above symbols denote time derivatives. 
Taking the positive scalar products of these
equations with the state $f\zo_0$, utilizing the orthonormalization
relations (\ref{orthon}), and solving for $\dot{a}_{n0}$, we find
\begin{mathletters}
\la{diffa}
\be
\dot{a}_{n0}=-\sum_j a_{nj}
e^{-\frac{i}{\hbar}\int_0^t E_j(t')dt'}{\angles{f\zo_0|\dot{f}\zo_j}}_+.
\ee
In a similar manner, we derive the equations
\be
\dot{a}_{nk}=-\sum_j a_{nj}
e^{-\frac{i}{\hbar}\int_0^t [E_j(t')-E_k(t')]dt'}
{\angles{f\zo_k|\dot{f}\zo_j}}_-
\ee
\end{mathletters}
for the coefficients corresponding to positive-norm states.

In order to estimate the decay of a state $f\zo_n$, we assume that at time
$t=0$ its expansion coefficients are $a_{nj}(0)=\delta_{nj}$. 
Approximating the slowly varying scalar products and energy eigenvalues
to be constant in time, and the decay to be negligible, such that we
may also treat the $a_{nj}$ coefficients on the rhs of Eqs.\ (\ref{diffa}) 
as constants, we can integrate them to yield
\be
\la{at}
a_{nk}(t)\simeq -\frac{i}{\omega_{nk}}
(e^{-i\omega_{nk} t} - 1){\angles{f\zo_k|\dot{f}\zo_n}}_{\pm}.
\ee
Above, the positive scalar product is chosen for $k=0$, the negative otherwise,
and we have denoted $\omega_{nk}=(E_n-E_k)/\hbar$. Since the 
quasi-stationary states are orthonormalized, the requirement of
negligible decay thus implies
\be
\la{gencriterion}
\bigg| \frac{1}{\omega_{nk}}{\angles{f\zo_k|\dot{f}\zo_n}}_{\pm} \bigg |
\ll 1.
\ee
Essentially, this is the validity criterion of the adiabatic approximation 
for the TDHFBP equations of the dilute boson gas.

Consider now the case of a vortex line precessing with frequency
$\nu_{\rm pr}$ about a circular orbit of radius $R_{\rm pr}$ 
in a harmonically trapped condensate;
for simplicity, we assume a trapping potential of the form
$V_{\rm tr}=\frac{1}{2}m\omega^2_r r^2$ in cylindrical coordinates
$\r=(r,\theta,z)$, and the vortex line to be directed along the $z$-axis.
In view of the differences in the vortex-core structure between
the BA and the self-consistent approximations, it is especially
interesting to find out whether the lowest-energy quasiparticles,
which constitute the major contribution to the noncondensate filling
the vortex core, can follow the moving vortex line rigidly, i.e., 
adiabatically. In order to assess the validity of the criterion 
(\ref{gencriterion}) for such states, we use the estimate
\be
\la{velappr}
{\angles{f\zo_k|\dot{f}\zo_n}}_{\pm}\simeq \vect{v}\cdot
{\angles{f\zo_k|\nabla f\zo_n}}_{\pm},
\ee
where $\vect{v}$ is the velocity of the vortex line. This approximation
treats accurately the region in the vicinity of the vortex line,
although it is exact only for a uniform vortex motion.
Furthermore, supposing the precession orbit is not too near the 
condensate boundary, we may use the quasiparticle states of a system with
a vortex located in the center of the trap to estimate the scalar products
on the rhs of Eq.\ (\ref{velappr}). Such a system is cylindrically symmetric, 
and the quasiparticle eigenstates can be chosen to be of the form
\begin{mathletters}
\label{eq:ansatz}
\begin{eqnarray}
u_q(\r)&=&{\rm u}_q(r)e^{iq_z(2\pi/L)z+i(q_{\theta}+1)\theta},\\
v_q(\r)&=&{\rm v}_q(r)e^{iq_z(2\pi/L)z+i(q_{\theta}-1)\theta},
\end{eqnarray}
\end{mathletters}
where $q_{\theta}$ and $q_z$ are integer angular and axial momentum 
quantum numbers, respectively, and $q$ denotes the complete set
of quantum numbers for the states. Calculation of the required matrix
elements is straightforward for these states---the result is
\ba
\la{finalint}
&&v I_{qq'}\equiv\vect{v}\cdot
{\angles{f\zo_q|\nabla f\zo_{q'}}}_{\pm}\nonumber\\
&&\simeq -\frac{v}{2}\delta_{q_z q_z'}
\delta_{|q_{\theta}-q'_{\theta}|,1}
\int_0^{\infty}dr\,\bigg[
r\parenth{{\rm u}^*_q\der{{\rm u}_{q'}}{r}
\pm{\rm v}^*_q\der{{\rm v}_{q'}}{r}}\nonumber\\
&&\qquad\qquad+(q'_{\theta}-q_{\theta})
[(q'_{\theta}+1){\rm u}^*_q {\rm u}_{q'}\pm
(q'_{\theta}-1){\rm v}^*_q {\rm v}_{q'}]\bigg],
\ea
where $v=|\vect{v}|$ is the magnitude of the 
velocity of the vortex line, and the
states are normalized according to 
\mbox{$\int_0^{\infty}rdr\,(|{\rm u}_q|^2\pm|{\rm v}_q|^2)=1$},
 with the
plus (minus) sign used for the zero-energy condensate state
(other states). Equations
(\ref{gencriterion}), (\ref{velappr}), and (\ref{finalint}) finally
yield the criterion
\be
\la{finalcrit}
v\ll v_{qq'}\equiv
\bigg|\frac{\omega_{qq'}}{I_{qq'}}\bigg|
\qquad\text{(for all $q$)}
\ee
for the velocity of the precessing vortex in order for the
state $f\zo_{q'}$ to follow the vortex rigidly. 

We have numerically computed the adiabaticity velocities $v_{qq'}$ 
for the lowest excitations of a cylindrical condensate. 
The static HFB equations
were solved self-consistently within the PA and its so-called G1 and
G2 variants \cite{hutchinson,proukakis}, 
in order to find the quasiparticle amplitudes
$u_q(\r)$, $v_q(\r)$, and the respective eigenenergies---for 
details of the methods used in the computations, see
Ref.\ \cite{our_letter}. In order to facilitate comparison with
the vortex precession observations, we use parameter values which
essentially correspond to the experiments reported in
Ref.\ \cite{new_exp}. Especially, the radial trapping frequency was set to
$\nu_r=\omega_r/2\pi=7.8$ Hz, and the density of the 
trapped $^{87}$Rb atoms was adjusted to yield a healing length 
$\xi=(8\pi n_0 a)^{-1/2}\approx 0.7$ $\mu$m at temperature 
$T\approx 0.8T_{\sc bec}$. Here $n_0$ denotes the peak density of the 
condensate, and the condensation temperature $T_{\sc bec}\approx 30$ nK.

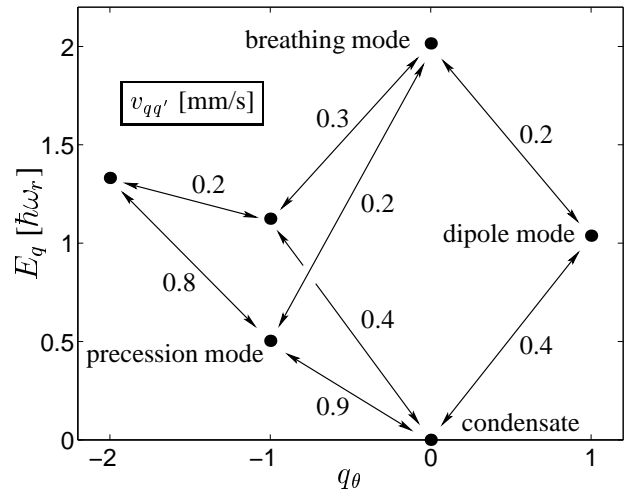
\begin{figure}
\caption{
The lowest vortex state collective modes with $q_z=0$,
and the adiabaticity velocities $v_{qq'}$ (see Eq.\ (\ref{finalcrit}))
determined by the transition rates between these states. Note especially
the low adiabaticity velocity given by the transitions between
the precession (LCLS) and breathing modes. The data corresponds
to the PA and system temperature $T\approx0.8T_{\sc bec}$.
}
\label{fig:1}
\end{figure}

In the experiments, the observed precession radii were of the order of
$R_{\rm pr}\simeq R/3\simeq 10$ $\mu$m, where $R$ denotes the radius
of the condensate \cite{new_exp}.
Bare-core vortices were observed to precess
in the direction of the condensate flow with frequency $\nu_r\approx 1.8$ Hz,
which corresponds to a velocity $v_{\rm exp}=2\pi R_{\rm pr} \nu_r
\approx 0.1$ mm/s. This is to be compared with the computed velocities
$v_{qq'}$ for the lowest quasiparticle states with $q_z=0$, displayed in Fig.\
\ref{fig:1} \cite{other_vqq}. Although $v_{\rm exp}\lesssim v_{qq'}$, we find
the adiabaticity condition (\ref{finalcrit}) not to be fulfilled. 
This suggests that, due to the deformation of the 
quasiparticle states, the noncondensate
can not follow the vortex line rigidly at these velocities. Especially 
interesting is the smallest adiabaticity velocity given by the decay of the 
so-called lowest core localized state (LCLS), which is the lowest
excitation with $(q_{\theta},q_z)=(-1,0)$, and itself corresponds to
the precession of the vortex. The LCLS has a crucial role in
the filling of the vortex core with noncondensate, which stabilizes
the static vortex state \cite{isoshima,our_letter}. In fact, this state is
almost solely responsible for the differences in the vortex structure
between the BA and PA in the low-temperature limit.
Deformation of the LCLS due to the vortex motion thus implies 
crucial modifications for the vortex core structure. 

The given adiabaticity velocities also hold for the G1 and G2, 
since differences between the approximations turn out to be negligible
in this respect. Computations with various parameter values also confirmed
the validity of the criterion (\ref{finalcrit}) to be 
largely independent of the specific values of  the
trapping frequency, the density of the gas, or the effective interaction
between the atoms. Essentially, the adiabaticity of the system
is determined by the precession radius $R_{\rm pr}$, via its
proportionality to the velocity of the precessing vortex line. In addition,
the adiabaticity could depend on the temperature: although we found
the smallest $v_{qq'}$ to depend only weakly on temperature, the stationary
PA predicts the precession mode frequency and, hence, the precession
velocity to have a strong temperature dependence \cite{our_letter}.

In conclusion, we have derived a criterion for the validity of the
quasi-stationary approximation for a time-dependent mean-field
formalism describing the dynamics of the condensate and thermal
components of a dilute boson gas. Application of this criterion
to a harmonically trapped Bose-Einstein condensate containing
an off-axis, precessing vortex line is shown to yield for the
vortex velocity a condition which is not fulfilled in the
experiments conducted so far. Deformation of the vortex structure
due to its motion is thus suggested to be at least partly responsible
for the apparent discrepancies between the predictions of the
stationary self-consistent approximations and the results of the
vortex precession experiments.

We thank the Center for Scientific Computing for computer resources,
and the Academy of Finland and the Graduate School in Technical Physics
for support.


\begin{references}

\bibitem{first_exp}
M.~H.~Anderson {\em et al.}, Science {\bf 269}, 198 (1995);
C.~C.~Bradley {\em et al.}, Phys.\ Rev.\ Lett.\ {\bf 75}, 1687 (1995);
K.~B.~Davis {\em et al.}, Phys.\ Rev.\ Lett.\ {\bf 75}, 3969 (1995).

\bibitem{vortex_exp}
M.~R.~Matthews {\em et al.},  Phys.\ Rev.\ Lett.\ {\bf 83}, 2498 (1999);
K.~W.~Madison {\em et al.}, Phys.\ Rev.\ Lett.\ {\bf 84}, 806 (2000).

\bibitem{fetter_review}
A.~L.~Fetter and A.~A.~Svidzinsky,
J.~Phys.: Condens.~Matter {\bf 13}, R135 (2001),
and references therein.

\bibitem{neg_modes}
R.~J.~Dodd {\em et al.},
Phys.\ Rev.\ A {\bf 56}, 587 (1997);
D.~S.~Rokhsar, Phys.\ Rev.\ Lett.\ {\bf 79}, 2164 (1997).

\bibitem{prec_lcls}
M.~Linn and A.~L.~Fetter,
Phys.\ Rev.\ A {\bf 61}, 063603 (2000);
A.~A.~Svidzinsky and A.~L.~Fetter,
Phys.\ Rev.\ Lett.\ {\bf 84}, 5919 (2000);
Phys.\ Rev.\ A {\bf 62}, 063617 (2000).

\bibitem{feder}
D.~L.~Feder {\em et al.}, Phys.\ Rev.\ Lett.\ {\bf 86}, 564 (2001).

\bibitem{new_exp}
B.~P.~Anderson {\em et al.},
Phys.\ Rev.\ Lett.\ {\bf 85}, 2857 (2000).

\bibitem{dalfovo_review}
F.~Dalfovo {\em et al.},
Rev.\ Mod.\ Phys.\ {\bf 71}, 463 (1999).

\bibitem{isoshima}
T.~Isoshima and K.~Machida, Phys.\ Rev.\ A {\bf 59}, 2203 (1999).

\bibitem{our_letter}
S.~M.~M.~Virtanen, T.~P.~Simula, and M.~M.~Salomaa,
Phys.\ Rev.\ Lett.\ {\bf 86}, 2704 (2001).

\bibitem{dodd}
R.~J.~Dodd {\em et al.}, Phys.\ Rev.\ A {\bf 57}, R32 (1998).

\bibitem{hutchinson}
D.~A.~W.~Hutchinson, R.~J.~Dodd, and K.~Burnett,
Phys.\ Rev.\ Lett.\ {\bf 81}, 2198 (1998).

\bibitem{tformalism}
A.~S.~Alexandrov and W.~H.~Beere, Phys.\ Rev.\ B {\bf 51}, 5887 (1995);
A.~Griffin, Phys.\ Rev.\ B {\bf 53}, 9341 (1996);
S.~Giorgini, Phys.\ Rev.\ A {\bf 57}, 2949 (1998).

\bibitem{schiff}
L.~I.~Schiff, {\em Quantum Mechanics} (McGraw-Hill, New York, 1955);
a similar analysis for vortices in superconductors has been presented in
E.~\v{S}im\'{a}nek, Phys.\ Rev.\ B {\bf 46}, 14054 (1992).

\bibitem{morgan}
S.~A.~Morgan,
J.\ Phys.\ B: At.\ Mol.\ Opt.\ Phys.\ {\bf 33}, 3847 (2000).

\bibitem{proukakis}
N.~P.~Proukakis {\em et al.}, 
Phys.\ Rev.\ A {\bf 58}, 2435 (1998).

\bibitem{other_vqq}
The adiabaticity velocities given by transitions between states with
higher energy or $q_z\neq 0$ were, in general, found to be larger.

\end{references}
\end{document}